\documentclass[12pt, a4paper]{article}

\usepackage[english]{babel}
\usepackage{graphicx} 
\usepackage{mathtools}
\usepackage{amsfonts}
\usepackage{enumitem}
\usepackage[colorlinks]{hyperref} 
\usepackage[affil-it,auth-sc]{authblk}

\title{Effects of different loading on the bifurcation of annular elastic rods: theory vs. experiments}

\author[1]{M. Gaibotti}
\author[1]{D. Bigoni\footnote{Corresponding author: e-mail: \href{mailto:bigoni@ing.unitn.it}{bigoni@ing.unitn.it}; phone: +39\,0461\,282507.}}
\author[2,3]{A. Cutolo}
\author[2,4]{M. Fraldi}
\author[1]{A. Piccolroaz}

\affil[1]{Instabilities Lab, University of Trento, Trento, Italy}
\affil[2]{Department of Structures for Engineering and Architecture, University of Napoli “Federico II”, Napoli, Italy}
\affil[3]{Laboratory of Integrated Mechanics and Imaging for Testing and Simulation (LIMITS), University of Napoli "Federico II", Napoli, Italy}
\affil[4]{LPENS - Départment de Physique, Ecole Normale Supérieure, Paris, France}

\date{Dedicated to Professor Giuseppe Saccomandi on the occasion of his 60th birthday}

\begin{document}
\maketitle
\begin{abstract}
\noindent 
The bifurcation problem of a circular Euler-Bernoulli rod subject to a uniform radial force distribution is investigated under three distinct loading conditions: (i.) hydrostatic pressure, (ii.) centrally-directed, and (iii.) dead load. Previous studies on this apparently \lq familiar' structural problem have yielded controversial results, necessitating a comprehensive clarification. This study shows that results previously labelled as \lq correct' or \lq wrong' simply refer to different external constraints, whose presence becomes necessary only for the two latter loads, (ii.) and (iii.). 
Moreover, the paper presents the first experimental realization of a circular rod subjected to centrally-directed loads. 
The experimental findings align with the theoretical predictions and show the exploitation of a new type of load acting on a continuous structural element. The feasibility of this load is demonstrated through the use of inextensible cables 
and opens the way to applications in flexible robotics when cables are used for actuation. 
\end{abstract}


\section{Introduction}

The in-plane bifurcation problem of circular elastic rods and arches, assumed axially-inextensible and loaded by hydrostatic pressure, is an old topic, which attracted considerable attention in civil and mechanical engineering (see the initial works by Bresse \cite{bresse1859cours} and Lévy \cite{levy1884memoire} and later among many others, \cite{tadjbakhsh1967equilibrium,oran1969buckling,wang1985symmetric,wang1993elastic,kang1994thin,pi2002plane,attard2013plane, wangcm}). 
Driven by new applications on minimal surfaces \cite{giomi} and the biology of several different natural structures \cite{catte, savin, napoli}, the issue has seen renewed interest.

Radial and uniform loads, leave an axially-inextensible circular rod undeformed and subject to a trivial state of pure normal compressive force until buckling occurs, usually in the form of an ovalization. 
However, initially identical load distributions may differ in the way they react to the deformation. In particular,   
hydrostatic pressure is just one type of {\it uniform and radial load} that a ring can experience. Specifically, the following three types of loads have been so far investigated for the circular rod \cite{boresi1955refinement, bodner1958conservativeness, armenakas1963vibrations}. 
\begin{itemize}
    \item[(i)] \textit{Hydrostatic pressure}, which remains orthogonal to the tangent to the deformed configuration of the rod. Moreover, the resultant force acting on the elementary arc of the rod changes proportionally to a variation in its length 
    (which cannot occur for axial inextensibility).
    \item [(ii)] \textit{Centrally-directed load}, which remains directed towards the initial centre of the ring. Moreover, the resultant force acting on the elementary arc of the rod is independent of a variation in its length. 
    This load can be visualized (and implemented in practice, as demonstrated in the present article) as several inextensible ropes pulling the rod and passing through a fixed point, coincident with the initial centre of the ring. 
    \item [(iii)] \textit{Dead load}, which remains directed along the normal to the rod in its undeformed configuration. Moreover, the resultant force acting on the elementary arc of the rod is independent of its deformation.
\end{itemize}
All loads (i.)--(iii.) become critical for buckling at a sufficiently high intensity, and infinite bifurcations arise at increasing values. The critical radial load $\Pi_{\text{cr}}$, corresponding to bifurcations, occurring in all possible modes and under every constraint externally applied to the rod, 
can, in any case, be expressed as \cite{singer1970buckling}
\begin{equation}
\label{eq:pcrk}
    \Pi_{\text{cr}}=k^2\frac{B}{R^3} ,
\end{equation}
where $R$ is the radius of the circle defining the undeformed configuration of the rod, $B=EJ$ its bending stiffness (equal to the product between the Young's modulus $E$ and the second moment of inertia of its cross-section $J$), and $k^2$ is a dimensionless constant depending on the type (i.)--(iii.) of radial load, on the selected mode of bifurcation, and on the constraints applied to the rod
(differences in  external constraints have been considered in \cite{armenakas1963vibrations,schmidt1980critical,schmidt1981buckling,mascolo2023revisitation}).  
In particular, the following values have previously been reported:
\begin{itemize}
    \item $k^2=3$ for hydrostatic pressure (i.) \cite{singer1970buckling}, 
    \item $k^2=9/2$ or $k^2 \approx 6.47$ for centrally-directed load (ii.)  \cite{boresi1955refinement}, 
    \item  $k^2 \approx 0.701$ or $k^2= 4$ for dead load (iii.) \cite{schmidt1980critical} .
\end{itemize}
The latter values (ii.) and (iii.) are controversial and are reported in the literature as \lq correct' or \lq wrong' \cite{schmidt1980critical,singer1970buckling}. 

\begin{figure}[hbt!]
    \centering
    \includegraphics[keepaspectratio, scale=1.16]{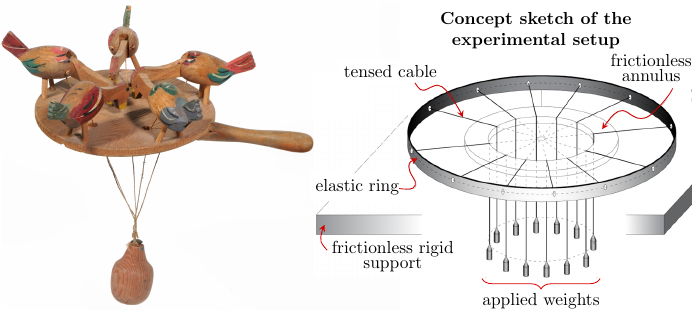}
    \caption{Left: An ancient children's toy, a sort of wooden automaton in which a small oscillating mass moves the hens to peck at the feed. The toy (Italian manufacture, early 20th century, courtesy of the F. Bigoni private collection) resembles the device on the right. Right: the basic idea behind the device's design for transmitting the centrally-directed load to the ring.}
    \label{uccelletti}
\end{figure}

The purpose of the present article is twofold: 
\begin{itemize}
    \item first to show that all the values for the buckling radial loads (ii.) and (iii.) so far presented are correct, but refer to different external constraints, imposed to prevent rigid-body displacements. In particular, while any system of statically-determined external constraints leaves the bifurcation problem under hydrostatic pressure (i.) unaffected, consideration of constraints becomes important for loads (ii.) and (iii.), because their application strongly changes the bifurcation loads and modes. Moreover, differently from centrally-directed load (ii.), the dead load (iii.) makes the structure unstable with respect to rigid-body rotations, so that in this case external constraints cannot be avoided;
    \item second, an experimental set-up is proposed to realize the load (ii.), showing that the experimental values of the critical load match with accuracy the theory. The realization of the centrally-directed load provides the design of a structure subject to a type of load proposed a long time ago and never achieved before. The device's scheme designed to reproduce the centrally-directed load is reported in Fig.~\ref{uccelletti}, together with an ancient toy based on a similar idea. 
\end{itemize}

The present article's results reiterate the importance of modelling external loads and clarify a controversial structural problem. Moreover, a new experimental strategy is introduced to attain centrally-directed loads. 
Though recently reconsidered \cite{hoang}, centrally-directed loads have been only scarcely analyzed, but they are of interest in the design of flexible robotic arms driven by cables, or pulley systems applied to deformable elements.

\section{Governing equations for the annular rod}

\begin{figure}[hbt!]
    \centering
    \includegraphics[keepaspectratio, scale=1.15]{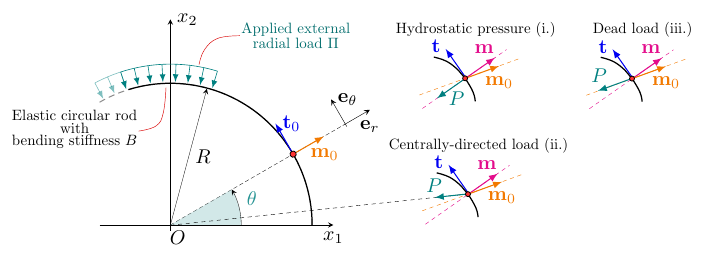}
    \caption{
    An elastic circular rod, centred in a Cartesian frame of reference $O-(x_1,x_2)$ and subjected to a radial load uniformly distributed $\Pi$. At every point of the rod, a tangent and a normal unit vectors, are defined, 
    in the undeformed and deformed configurations, 
    $\mathbf{t}_0$, $\mathbf{m}_0$, 
    and
    $\mathbf{t}$, $\mathbf{m}$,
        together with a local polar coordinate system with unit tangent and normal vectors  $\mathbf{e}_\theta$ and $\mathbf{e}_r$. Depending on the direction that the resultant force $P$ of the applied radial load on a deformed element assumes in the deformed configuration it is possible to define (i.) a \textit{hydrostatic load}, where $P$ is directed along $\mathbf{m}$, (ii.) a \textit{centrally-directed load} where  $P$ points towards point $O$, and (iii) a \textit{dead load} where $P$ is directed along $\mathbf{m}_0$
        }
    \label{fig:reference}
\end{figure}

Consider an inextensible and unshearable circular rod, characterized by a radius $R$, a bending stiffness $B$, and a Cartesian reference system with axes $x_1$ and $x_2$ centred at the centre $O$ of the structure. The arc length $ds=Rd\theta$ is defined with respect to a polar coordinate system ($r, \theta$). At every point of the rod, a tangential $\mathbf{t}_0$ and a radial $\mathbf{m}_0=\mathbf{t}_0\times{\mathbf{e}_3}$ unit vectors are introduced, where $\mathbf{e}_3$ is the out-of-plane unit vector,  Fig. \ref{fig:reference}. 
%
%
In the Cartesian frame of reference $(x_1,x_2)$ described by the unit vectors $\mathbf{e}_1$ and $\mathbf{e}_2$, the tangent and the normal unit vectors at a point on the rod assumes the form
\begin{equation}
\label{eq:tan_norm}
    \mathbf{t}_0=-\sin \theta \mathbf{e}_1+\cos \theta \mathbf{e}_2 , 
    \quad 
    \mathbf{m}_0=\cos\theta \mathbf{e}_1+\sin \theta \mathbf{e}_2 .
\end{equation}
The displacement vector describing points belonging to the rod is 
\begin{equation}
\label{eq:displacement}
    \mathbf{u}=u_{\theta}\,\mathbf{t}_0+u_{r}\,\mathbf{m}_0 ,
\end{equation}
where $u_r$ and $u_\theta$ are the radial and tangential components, with respect to the orthogonal unit vectors
\begin{equation}
    \mathbf{e}_r=\cos\theta\,\mathbf{e}_1 + \sin\theta\,\mathbf{e}_2,  
    \quad
    \mathbf{e}_\theta=-\sin\theta\,\mathbf{e}_1 + \cos\theta\,\mathbf{e}_2,
\end{equation}
which define the radial and circumferential directions.

The axial deformation $\epsilon$, the cross-section rotation $\Phi$ and the change of curvature $\chi$ at every point of the rod are governed by \cite{timoshenko1950strength}
\begin{equation}
\label{eq:kinematics}
    \begin{aligned}
        \epsilon=\frac{u_r}{R}+\frac{\partial{u_\theta}}{\partial{s}}, && \Phi=\frac{\partial{u_r}}{\partial{s}}-\frac{u_{\theta}}{R}, && \chi=-\frac{\partial{\Phi}}{\partial{s}},
    \end{aligned}
\end{equation}
respectively. Assuming the inextensibility of the rod, $\epsilon=0$, and introducing the constitutive equation for in-plane deflection, it follows from \eqref{eq:kinematics}$_1$
\begin{equation}
\label{eq:inextensibility}
    \begin{aligned}
        u_r=-\frac{\partial{u_\theta}}{\partial{\theta}}, && \chi=\frac{M}{B} , 
    \end{aligned}
\end{equation}
where $M$ is the bending moment internal to the rod. When external pressure is applied, the ring is only subject to a uniform internal compressive force $N_0=-\Pi R$, while both the shearing force $T_0$ and the bending moment $M_0$ are null. 
The equilibrium equations of any curved rod (not necessarily circular), subject to the load $\mathbf{q}$, are (see details in \cite{gaibotti2024bifurcations})
\begin{equation}
\label{eq:equil_B0}
    \begin{array}{ll}
    &\dfrac{\partial{N_{0}}}{\partial{s}}+\dfrac{T_0}{R}=-\mathbf{q} \cdot \mathbf{t}_0, \quad
    \dfrac{N_0}{R}-\dfrac{\partial{T_{0}}}{\partial{s}}=\mathbf{q} \cdot \mathbf{m}_0,\\[4 mm] 
    &\dfrac{\partial{M_{0}}}{\partial{s}}=T_{0}\left(\dfrac{u_{r}}{R}+\dfrac{\partial{u_{\theta}}}{\partial{s}}+1\right)-N_{0}\left(\dfrac{\partial{u_{r}}}{\partial{s}}-\dfrac{u_{\theta}}{R}\right) ,
    \end{array}
\end{equation}
so that, assuming the curved configuration as reference in a relative Lagrangian description ($u_r=u_\theta=0$), the material time derivative leads to the incremental equilibrium equations
\begin{equation}
\label{eq:equil_incr}
    \dfrac{\partial{\dot{N}_{0}}}{\partial{s}}+\dfrac{\dot{T}_{0}}{R}=-\dot{\mathbf{q}}\cdot{\mathbf{t}_{0}} , 
    \quad
    \dfrac{\dot{N}_{0}}{R}-\dfrac{\partial{\dot{T}_{0}}}{\partial{s}}=\dot{\mathbf{q}}\cdot{\mathbf{m}_{0}} ,
    \quad
    \dfrac{\partial{\dot{M}_0}}{\partial{s}}=\dot{T}_{0}+\Pi R\left(\dfrac{\partial{\dot{u}_{r}}}{\partial{s}}-\dfrac{\dot{u}_{\theta}}{R}\right) ,
\end{equation}
where a superimposed dot denotes an increment, while the load increment, $\dot{\mathbf{q}}$, depends on the type of load (i)-(iii). The material time derivative of equation \eqref{eq:inextensibility}$_2$ and the use of the relation \eqref{eq:kinematics}$_{2-3}$ yield 
\begin{equation}
\label{eq:Mincr}
	\frac{\partial{\dot{M}_{0}}}{\partial{s}}=-B\left(\frac{\partial^{3}{\dot{u}_{r}}}{\partial{s^{3}}}+\frac{\partial{\dot{u}_{r}}}{\partial{s}}\frac{1}{R^{2}}\right) .
\end{equation}
Therefore, a substitution of equation \eqref{eq:Mincr} into  equation \eqref{eq:equil_incr}$_3$, allows to reduce all equations \eqref{eq:equil_incr} into one equation describing the incremental response of a circular rod \cite{gaibotti2024bifurcations} as 
\begin{equation}
\label{eq:govIncr_gen}
    \begin{aligned}
        & \frac{\partial^{6}{\dot{u}_{\theta}}}{\partial{\theta^{6}}}+\left(2+k^2\right)\frac{\partial^{4}{\dot{u}_{\theta}}}{\partial{\theta^{4}}}+\left(1+2k^2\right)\frac{\partial^2{\dot{u}_{\theta}}}{\partial{\theta}^2}+k^2\dot{u}_\theta+\mathfrak{S}=0 , \\
        & \dot{u}_{r} + \frac{\partial \dot{u}_\theta}{\partial{\theta}} = 0 ,
    \end{aligned}
\end{equation}
where
\begin{equation}
\label{eq:S}
    \mathfrak{S}=\frac{R^4}{B}\left(\frac{\partial\dot{\mathbf{q}}}{\partial\theta}\cdot{\mathbf{m}}_0+2\dot{\mathbf{q}}\cdot\mathbf{t}_0\right) .
\end{equation}
The incremental load $\dot{\mathbf{q}}$ in equation \eqref{eq:S} is one of the incremental loads corresponding to (i)--(iii). These can be written as the equations (3.18)-(3.20)$_2$ derived and reported in \cite{gaibotti2024bifurcations} and leading to 
\begin{equation}
\label{eq:dotPi}
    \dot{\mathbf{q}} = 
    -\frac{\Pi}{R} \times 
    \left\{
    \begin{array}{ll}
        \left(\dfrac{\partial^{2}{\dot{u}_{\theta}}}{\partial{\theta^2}} + \dot{u}_{\theta}\right)\mathbf{t}_{0} & \text{for hydrostatic pressure (i.)} , \\[3mm]
        \dot{u}_\theta\,\mathbf{t}_0 & \text{for 
         centrally-directed load (ii.)} , \\[3mm]
        \bf{0} & \text{for dead load (iii.)} .
    \end{array}
    \right.
\end{equation}
A substitution of equations \eqref{eq:dotPi} into the equation \eqref{eq:S} yields 
\begin{equation}
\label{eq:sTerm}
    \mathfrak{S}=-k^2\,\times\left\{
    \begin{array}{lll}
         \dfrac{\partial^2\dot{u}_\theta}{\partial \theta^2}+\dot{u}_\theta & \text{for hydrostatic pressure (i.)} , \\[3mm]
         \dot{u}_\theta & \text{for 
         centrally-directed load (ii.)} , \\[3mm]
         0 & \text{for dead load (iii.)} .
    \end{array}\right.
\end{equation}

The internal axial $\dot{N}_0$ and tangential $\dot{T}_0$ incremental forces and bending moment $\dot{M}_0$ in equations \eqref{eq:govIncr_gen} have the following form
\begin{equation}
\label{eq:ntm}
    \begin{aligned}
        &\dot{N}_0=\frac{B}{R^3}\left(\frac{\partial^5\dot{u}_\theta}{\partial \theta^5}+\frac{\partial^3\dot{u}_\theta}{\partial \theta^3}\right)+\Pi\left(\frac{\partial^3\dot{u}_\theta}{\partial\theta^3}+\frac{\partial\dot{u}_\theta}{\partial \theta}\right), \\
        &\dot{T}_0=\frac{B}{R^3}\left(\frac{\partial^4 \dot{u}_\theta}{\partial \theta^4}+\frac{\partial^2 \dot{u}_\theta}{\partial\theta^2}\right)+\Pi\left(\frac{\partial^2\dot{u}_\theta}{\partial \theta^2}+\dot{u}_\theta\right) ,\\
        &\dot{M}_0=\frac{B}{R^2}\left(\frac{\partial^3 \dot{u}_\theta}{\partial \theta^3} + \frac{\partial\dot{u}_\theta}{\partial \theta}\right) .
    \end{aligned}
\end{equation}

\section{Bifurcation analysis}

Depending on the behaviour of the externally applied radial load during the deformation, \cite{bodner1958conservativeness,armenakas1963vibrations} the following cases have to be analyzed.
\begin{itemize}
    \item[(i)] For hydrostatic pressure, the governing equation is\cite{singer1970buckling}
    \begin{equation}
    \label{eq:goveqHydro}
        \frac{\partial^{6}\dot{u}_{\theta}}{\partial{\theta^6}}+(2+k^2)\frac{\partial^{4}\dot{u}_{\theta}}{\partial{\theta^4}}+(1+k^2)\frac{\partial^{2}\dot{u}_{\theta}}{\partial{\theta^2}}=0 ,
    \end{equation}
    and its general solution can be written as 
    \begin{equation}
    \label{eq:genHydro}
        \dot{u}_{\theta}(\theta)=a_1+b_1 \theta+a_2 \cos \theta+a_3 \sin \theta+b_2 \cos\omega\theta+b_3 \sin\omega\theta ,
    \end{equation}
    where $a_1$--$a_3$ and $b_1$--$b_3$ are integration constants and
    \begin{equation}
    \label{eq:alpha12}
        \omega=\sqrt{k^2+1}.
    \end{equation}
    \item[(ii)] For centrally-directed load, the governing equation is \cite{singer1970buckling}
    \begin{equation}
    \label{eq:goveqCentrally}
        \frac{\partial^{6}\dot{u}_{\theta}}{\partial{\theta^6}}+(2+k^2)\frac{\partial^{4}\dot{u}_{\theta}}{\partial{\theta^4}}+(1+2k^2)\frac{\partial^{2}\dot{u}_{\theta}}{\partial{\theta^2}}=0 ,
    \end{equation}
    and its general solution can be written as 
    \begin{equation}
    \label{eq:genCentr}
        \dot{u}_{\theta}(\theta)=a_1+b_1 \theta+b_2 \cos\omega_1\theta +b_3 \sin\omega_1\theta +b_4 \cos\omega_2\theta+b_5 \sin\omega_2\theta ,
    \end{equation}
    where $a_1$ and $b_1$--$b_5$ are integration constants and 
    \begin{equation}
    \label{eq:genCentr_omega}
    \begin{aligned}
        \omega_1=\sqrt{1+\frac{k}{2}  (k+\sqrt{k^2-4})}, && \omega_2=\sqrt{1+\frac{k}{2}  (k-\sqrt{k^2-4})} .    \end{aligned}
    \end{equation}
    Note that for a fixed value of $\omega_1=\omega_1^0>\sqrt{3}$, eq.~(\ref{eq:genCentr_omega})$_1$ has only a unique solution $k^0$ for $k$. Then $-k^0$ solves eq.~(\ref{eq:genCentr_omega})$_2$ for $\omega_2=\omega_1^0$. Finally, $\omega_1=\omega_2$ when $k^2=4$. In this particular case, the solution of the differential equation becomes 
    \begin{equation}
    \label{eq:genCentr44}
        \dot{u}_{\theta}(\theta)=a_1+b_1 \theta+\left(b_2+b_3 \theta\right)\cos \sqrt{3} \theta +\left(b_4 + b_5 \theta\right)\sin\sqrt{3} \theta ,
    \end{equation}
    \item[(iii)] For dead load, the governing equation is \cite{singer1970buckling}
    \begin{equation}
    \label{eq:goveqConst}
        \frac{\partial^{6}\dot{u}_{\theta}}{\partial{\theta^6}}+(2+k^2)\frac{\partial^{4}\dot{u}_{\theta}}{\partial{\theta^4}}+(1+2k^2)\frac{\partial^{2}\dot{u}_{\theta}}{\partial{\theta^2}}+k^2\dot{u}_{\theta}=0 .
    \end{equation}
    and its general solution can be written as 
    \begin{equation}
    \label{eq:genConst}
        \dot{u}_\theta(\theta)=a_2 \cos \theta+a_3 \sin \theta+b_1 \cos k\theta+b_2 \sin k\theta+b_3 \theta \cos \theta+b_4 \theta \sin \theta ,
    \end{equation}
    where $a_2$--$a_3$ and $b_1$--$b_4$ are integration constants. Note that in the particular case $k=1$, the solution becomes 
    \begin{equation}
    \label{solettona}
        \dot{u}_{\theta}(\theta)=a_2\cos\theta+a_3\sin\theta+(b_1 \theta^2+b_2 \theta)\cos\theta +(b_3 \theta^2+b_4 \theta)\sin \theta .
    \end{equation}
\end{itemize}

\subsection{Effect of the boundary conditions on the bifurcation}

Boundary conditions are to be imposed on solutions \eqref{eq:genHydro}, \eqref{eq:genCentr}, and \eqref{eq:genConst}.

\subsubsection{The role of rigid-body roto-translations on the equilibrium of the circular rod}
\label{ocadura}

As it has been so far presented, the circular rod is free in the plane and can suffer, in principle, a rigid-body roto-translation. 
This displacement is governed by the constants $a_1$, $a_2$, and $a_3$ in equations \eqref{eq:genHydro}, \eqref{eq:genCentr}, \eqref{eq:genConst} and can be represented as
\begin{equation}
\label{rgb_ring}
    \begin{aligned}
        &\dot{u}_{\theta}=a_1+a_2\cos\theta+a_3\sin\theta ,\\
        &\dot{u}_{r}=a_2\sin\theta-a_3\cos\theta ,
    \end{aligned}
\end{equation}
where $a_1$ corresponds to a rigid-body rotation, while $a_2$ and $a_3$ rule the vertical and horizontal rigid-body translations, respectively. 
However, not all the rigid-body displacements are compatible with the applied radial loads (i.)--(iii.), so that in some cases, work is produced during the rigid-body displacements. 

\paragraph{(i.) For hydrostatic pressure}
    all rigid-body displacements do not produce any work (for the undeformed, but also for an arbitrarily deformed, configuration of structure), so that 
    the expressions \eqref{rgb_ring} trivially satisfy the governing equation \eqref{eq:goveqHydro}. Therefore, in the bifurcation problem, constants $a_1$, $a_2$, and $a_3$ remain arbitrary in equation \eqref{eq:genHydro} and any (strictly necessary) external constraint system, which eliminates rigid body motions (for instance a clamp or three rollers), can be applied without changing the bifurcation loads and modes. 

\begin{figure}[hbt!]
	\centering\includegraphics[keepaspectratio, scale=1.15]{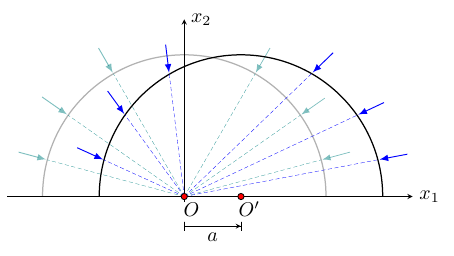}
    \caption{A rigid-body translation from point $O$ to point $O'$ of an annular rod (one half is reported), loaded with a centrally-directed pressure, breaks equilibrium. The symmetric configuration (shown grey in the background) is stable for a compressive radial load, so that when displaced, the structure spontaneously returns to the original configuration.}
    \label{rgbm_centrally}
    \end{figure}

\paragraph{(ii.) For centrally-directed radial load} 
    only the rigid-body rotation $a_1$ does not produce work, trivially satisfying equation \eqref{eq:goveqCentrally}, and thus remains undetermined in the incremental problem, equation (\ref{eq:genCentr}). However, it will be shown below that rigid-body translations always produce negative work, so that the structure will not move, even without constraints. The latter condition is compatible with certain external constraints (for instance three axial rollers inclined at angles $0$, $\pi/2$ and $\pi$). In this way, $k^2=9/2$ is obtained. If the external constraints are changed, for instance introducing a clamp, certain bifurcation modes are excluded and the bifurcation load increases at $k^2 \approx 6.47$. When a rigid-body translation is applied, Fig.~\ref{rgbm_centrally}, the centrally-directed load performs a non-null work.
    In particular, a rigid-body translation of a finite amount $a<R$ is postulated for the ring, aligned parallel to the horizontal axis $x_1$, so that the centre of the circular rod is displaced from $O$ to $O^{\prime}$. After this displacement, the resultant $d\mathbf{f}$ of the radial force $\Pi$ applied on an elementary arch of length $ds=Rd\theta$ is 
    \begin{equation}
	\label{eq:centrally_resultantB}
		d\mathbf{f}=-\Pi\frac{\left(\cos\theta+a/R\right)\mathbf{e}_1+\sin\theta\,\mathbf{e}_2}{\sqrt{a^2/R^2+2a/R\cos\theta+1}}ds.
	\end{equation}
    The work $W(a)$ done by the centrally-directed load during the application of the rigid-body translation of amount $a$ is obtained through a double integration of the scalar product of equation (\ref{eq:centrally_resultantB}) with $\mathbf{e}_1$ as
	\begin{equation}
	\label{eq:centrally_work}
        W(a)=-\Pi R^2 \int_{0}^{a/R}\left(\int_{0}^{2\pi}\frac{\cos\theta+\alpha}{\sqrt{\alpha^2+2\alpha\cos\theta+1}}\,d\theta \right) \, d\alpha .
	\end{equation}
    Recalling that $a < R$, the sign of the work may be estimated by considering the bounds
	\begin{equation}
	\label{cimancava2}
        \int_{0}^{\pi}\frac{\cos\theta}{1+\alpha}\,d\theta
        +
        \int_{\pi}^{2\pi}\frac{\cos\theta}{1-\alpha}\,d\theta
        +
        \frac{2\pi\alpha}{1+\alpha}
        \leq
        \int_{0}^{2\pi}\frac{\cos\theta+\alpha}{\sqrt{\alpha^2+2\alpha\cos\theta+1}}\,d\theta ,
	\end{equation}
    and
    \begin{equation}
	\label{cimancava4}
        \int_{0}^{\pi}\frac{\cos\theta}{1-\alpha}\,d\theta
        +
        \int_{\pi}^{2\pi}\frac{\cos\theta}{1+\alpha}\,d\theta
        +
        \frac{2\pi\alpha}{1-\alpha}
        \geq
        \int_{0}^{2\pi}\frac{\cos\theta+\alpha}{\sqrt{\alpha^2+2\alpha\cos\theta+1}}\,d\theta ,
	\end{equation}
    so that eventually
    \begin{equation}
	\label{cimancava6}
        0<\frac{2\pi\alpha}{1+\alpha}
        \leq
        \int_{0}^{2\pi}\frac{\cos\theta+\alpha}{\sqrt{\alpha^2+2\alpha\cos\theta+1}}\,d\theta 
        \leq \frac{2\pi\alpha}{1-\alpha} ,
	\end{equation}
    and therefore
    \begin{equation}
    \label{baldracca}
        a/R+\log(1-a/R) \leq \frac{W(a)}{2\pi R^2\Pi} \leq -a/R+\log(1+a/R) <0.
    \end{equation}
    It follows from the bounds \eqref{baldracca} that the work is always negative for compressive radial forces. It can be concluded that for compressive (for tensile) centrally-directed radial load, $\Pi>0$ ($\Pi<0$), the ring is stable (is unstable) to rigid-body translations, so that experiments on the ring are possible for $\Pi>0$ even without external constraints. 
 
\begin{figure}[hbt!]
	\centering
    \includegraphics[keepaspectratio, scale=1.15]{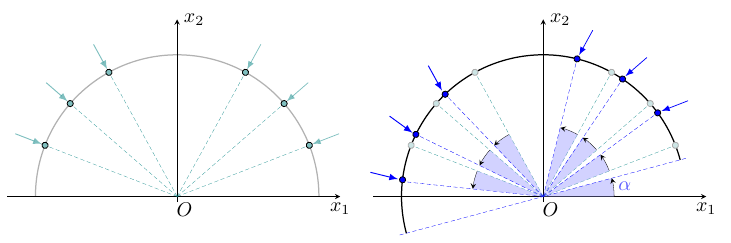} 
    \caption{A rigid-body rotation about the centre $O$ of an annular rod (one half is reported) subject to a dead radial load breaks equilibrium. The symmetric configuration (left) is unstable upon rotations (right).}
	\label{rgbm_constant}
    \end{figure}

\paragraph{(iii.) For dead radial load}  
    only the rigid-body translations $a_2$ and $a_3$ do not produce work, trivially satisfying equation \eqref{eq:goveqConst}, and therefore remain undetermined in the incremental problem, equation (\ref{eq:genConst}). 
    It will be shown below that any rigid-body rotation always produces positive work for compressive radial load, so that the structure will move and this movement has to be eliminated with a constraint. The latter condition has to leave unaffected the involved bifurcation mode, so that the first bifurcation mode is obtained with a clamp, $k^2\approx 0.701$, while three axial rollers determine $k^2=4$. When a finite rigid-body rotation $\alpha$ is applied to the annular rod, Fig.~\ref{rgbm_constant}, every point of its axis (determined by the angle $\theta$) suffers  the finite displacement $\mathbf{u}$ 
    \begin{equation}
    \label{eq:disp_DLoad}
        \mathbf{u}(\theta, \alpha)=-R(1-\cos\alpha)\,\mathbf{e}_r(\theta) + R\sin\alpha\,\mathbf{e}_\theta(\theta) .
    \end{equation}
    The resultant $d\mathbf{f}$ of the radial force $\Pi$ applied on an elementary arch of length $ds$ is 
    \begin{equation}
    \label{eq:f_DLoad}
        d\mathbf{f}=-\Pi ds\, \mathbf{e}_r ,
    \end{equation}	
    thus the work done by the whole dead radial load associated with the rotation $\alpha$ becomes
	\begin{equation}
	\label{eq:work:DLoad}
		-\Pi R\int_{0}^{2\pi}{\mathbf{e}_r\cdot\mathbf{u}(\theta,\alpha)\,d\theta}=2\pi R^2\Pi(1-\cos\alpha) .
	\end{equation}
	It follows from equation \eqref{eq:work:DLoad} that the work is always positive for the compressive radial load (or null in the trivial case $\alpha=2\pi$). It can be concluded that for compressive (tensile) dead radial load, $\Pi>0$ ($\Pi<0$), the ring is unstable (stable) to rigid-body rotations, in analogy to a rigid rod subject to two equal and opposite dead forces at its ends.    

\subsubsection{Circular rod: fully continuous bifurcation modes} 
\label{selfEq_sect}

Solutions (\ref{eq:genHydro}), (\ref{eq:genCentr}), and (\ref{eq:genConst}) and their derivatives are continuous functions of $\theta \in [0, 2\pi]$, so that continuity of the structural element is enforced by requiring that the function assumed the same value in 0 and in $2\pi$. In this Section, solutions are sought that respect the continuity of the incremental kinematic descriptors $\dot{u}_\theta$, $\dot{u}_r$, $\dot{\Phi}$ 
\begin{equation}
\label{eq:continuity}
    \dot{u}_\theta(0)=\dot{u}_\theta(2\pi), \quad \dot{u}_r(0)=\dot{u}_r(2\pi), \quad \dot{\Phi}(0)=\dot{\Phi}(2\pi), 
\end{equation}
and of the incremental internal forces, $\dot{M}$, $\dot{T}$, and $\dot{N}$
\begin{equation}
\label{eq:continuity2}
    \dot{M}(0)=\dot{M}(2\pi), \quad \dot{T}(0)=\dot{T}(2\pi), \quad \dot{N}(0)=\dot{N}(2\pi). 
\end{equation}
Therefore, an application eqs.~(\ref{eq:kinematics}) and (\ref{eq:ntm}), shows that continuity equations (\ref{eq:continuity}) and (\ref{eq:continuity2}) become equivalent to 
\begin{equation}
\label{palline}
    \frac{\partial^n \dot{u}_\theta}{\partial \theta^n}(0) =
    \frac{\partial^n \dot{u}_\theta}{\partial \theta^n}(2 \pi), \quad 
    n=0,...,5,
\end{equation}
where $n=0,1,2$ for the continuity of the kinematic descriptors and $n=3,4,5$ for the internal forces. 

The solutions (\ref{eq:genHydro})--(\ref{eq:genConst}) show that, when present, all coefficients $a_1$, $a_2$, and $a_3$ remain unaffected by the continuity conditions  (\ref{palline}), because they represent rigid-body motions, which a-priory satisfy the continuity of any order. Therefore, only a limited number of eqs.~(\ref{palline}) are to be used, in particular, six conditions minus the number of constants $a_i$. The conditions which are not imposed are automatically satisfied. 
\begin{itemize}
    \item[(i)] For hydrostatic pressure, equation (\ref{eq:genHydro}) shows that $b_1=0$ and that 
    \begin{equation}
    \label{peracotta}
        \left[
        \begin{array}{cc}
            \cos 2\pi\omega  -1 &  \sin 2\pi\omega \\
            \sin 2\pi\omega &  -\cos 2\pi\omega +1\\ 
            \end{array}
        \right]
        \left[
        \begin{array}{cc}
            b_2  \\
            b_3 
        \end{array}    
        \right]=0,
    \end{equation}
    so that non-trivial solutions may exist when 
    \begin{equation}
    \label{figa}
        \sin^2 \omega \pi = 0, ~~~ \Longrightarrow ~~~ \omega \mbox{ integer}.
    \end{equation}
    When $\omega$ is an integer, all the items in the matrix (\ref{peracotta}) vanish, so that the constants $b_2$ and $b_3$ remain undetermined. 
    Therefore, at bifurcation, $a_1$, $a_2$, $a_3$, $b_2$, and $b_3$ are all left  arbitrary by the conditions of continuity (\ref{palline}). The bifurcation modes, eq.~(\ref{eq:genHydro}), become
    \begin{equation}
    \label{azzzzz}
        \dot{u}_{\theta}(\theta)=a_1+a_2 \cos \theta+a_3 \sin \theta+b_2 \cos \theta\omega+b_3 \sin \theta\omega .
    \end{equation}
    Note that $\omega=1$ is a solution of equation (\ref{figa}) leading to $k=0$, a trivial condition which has to be disregarded, because it corresponds to rigid-body displacements. Therefore, the smallest value of critical load can be obtained from equation (\ref{figa}) as $\omega=2$, leading to $k^2=3$. 
    \item[(ii)] For centrally-directed load, equation (\ref{eq:genCentr}) shows that continuity requires $b_1=0$. In addition, the continuity of $\dot{u}_\theta$ up to its fifth derivative leads to an eigenvalue problem becoming singular when one of two independent conditions similar to eq.~(\ref{peracotta}) are satisfied, one involving $b_2$ and $b_3$ and the other $b_4$ and $b_5$, these respectively are 
    \begin{equation}
    \label{duepa}
        \sin^2 \omega_1 \pi = 0, \quad \text{ or } \quad \sin^2 \omega_2 \pi = 0,
    \end{equation}
    leading to integer values of $\omega_1$ and $\omega_2$. The two conditions (\ref{duepa}) are equivalent, so that bifurcation can be reduced to the request that $\omega_1$  be an integer and the bifurcation modes, eq. (\ref{eq:genCentr}), becomes  
    \begin{equation}
    \label{eq:genCentr33}
        \dot{u}_{\theta}(\theta)=a_1+b_2 \cos \theta \omega_1+b_3 \sin \theta \omega_1 .
    \end{equation}
    Note that the solutions $\omega_1 =1$ and $\omega_2=1$ of equations (\ref{duepa}) are to be disregarded as they lead to $k=0$, corresponding to a trivial bifurcation characterized by a rigid-body rotation governed by the arbitrary coefficient $a_1$. Additionally, the case $k^2=4$ corresponds to $\omega_1=\omega_2$, thus the corresponding general solution is given by eqn.~(\ref{eq:genCentr44}), which is not compatible with the required continuity conditions \eqref{palline}. The smallest value of critical load can be obtained from equation (\ref{duepa}$_1$) for $\omega_1=2$, leading to $k^2=9/2$. 
    \item[(iii)] For dead load, equation (\ref{eq:genConst}) shows that $b_3=b_4=0$, while 
    \begin{equation}
    \label{duepall}
        \sin^2 k \pi = 0 ,
    \end{equation}
    leading to integer values for $k$. Therefore, at bifurcation load $b_3=b_4=0$, while $a_2$, $a_3$, $b_1$, and $b_2$ remain unprescribed. The bifurcation modes, eq.~(\ref{eq:genConst}), become  
    \begin{equation}
    \label{eq:genConst2}
        \dot{u}_\theta(\theta)=a_2 \cos \theta+a_3 \sin \theta+b_1 \cos k\theta+b_2 \sin k\theta .
    \end{equation}
    Note that the solution $k =1$ of equation (\ref{duepall}) is to be disregarded, because eqn.~(\ref{solettona}) does not admit continuous solutions. As a conclusion, the smallest value for the critical load can be obtained from equation (\ref{duepall}) as $k^2=4$.
\end{itemize}

\begin{figure}[hbt!]
    \centering
    \includegraphics[keepaspectratio, scale=1.0]{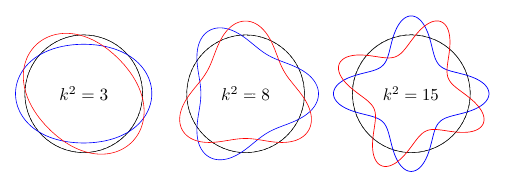}
    \includegraphics[keepaspectratio, scale=1.0]{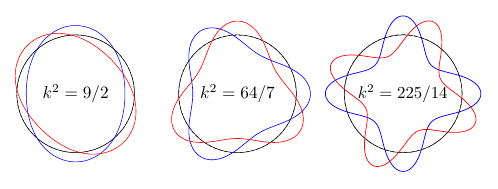}
    \includegraphics[keepaspectratio, scale=1.0]{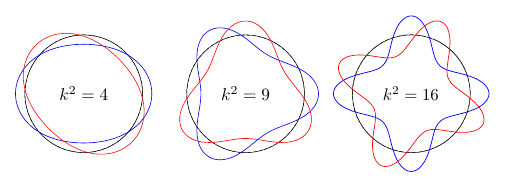}
    \caption{The first three bifurcation modes for (i.) hydrostatic pressure (upper row), (ii.) centrally-directed load (middle row) and (iii.) dead load (lower row), when the full continuity expressed by the relations \eqref{eq:continuity} and \eqref{eq:continuity2} is enforced. Two independent modes always occur, one sketched blue and the other red.}
    \label{hydro_continuity_ovetti}
\end{figure}

The first three bifurcation modes corresponding to the above \lq fully-continuous' solutions are reported in Fig.~\ref{hydro_continuity_ovetti}, for all types of investigated loads.

All the bifurcation modes shown in the figure are double, so that one is depicted as blue and the other red. It should also be noted that the first mode of bifurcation can be obtained without external constraints only in the cases of hydrostatic pressure and centrally-directed loaded. The first mode for the dead load cannot be realized without a strong external constraint system, as detailed in the next section.

\subsection{External constraints}

In the presence of external constraints, the solutions corresponding to fully continuous bifurcation's modes may no longer be valid. In fact, constraints introduce discontinuities; for instance, at a clamp, all the internal forces and moments may jump. When external constraints are present, the solutions (\ref{eq:genHydro}), (\ref{eq:genCentr}), and (\ref{eq:genConst}) are valid only within the intervals of $\theta$ comprised between each pair of constraints, so that six integration constants are to be obtained for each interval, by imposing the relevant conditions. For instance, a pin enforces the displacement components to vanish for both connected intervals (four conditions), plus the continuity of rotation and bending moment (two conditions). In the following, the possibility of achieving a fully continuous bifurcation solution is scrutinized with a view to external constraints. 

\subsubsection{(i)  Hydrostatic pressure}

For hydrostatic pressure, the fully continuous solution (\ref{eq:genHydro}) contains all the rigid-body displacement components, constants $a_1$, $a_2$, and $a_3$. Therefore, any well-assigned system of external constraints, which is statically determinate, is compatible with all fully continuous bifurcation modes. For instance, three rollers, or two rollers and a pin, or a clamp, are all possible external constraints compatible with the attainment of all fully continuous bifurcation modes. 
In particular, the first mode becomes visible, while the attainment of higher-order modes requires the use of statically-indeterminate external constraints, selected in a proper way. However, the equilibrium neutrality of every possible deformed shape of the ring under pressure loading, implies that the first bifurcation load and mode can be obtained even in the absence of external constraints (for instance depressurizing a tube, Fig.~1 of \cite{gaibotti2024bifurcations}). 

\subsubsection{(ii) Centrally-directed load} 

\begin{figure}[hbt!]
    \centering    
    \includegraphics[keepaspectratio, scale=1.0]{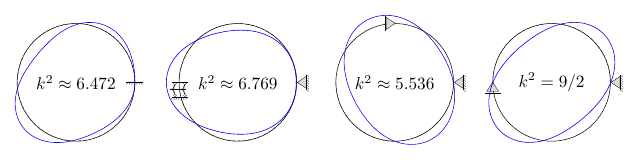}
    \includegraphics[keepaspectratio, scale=1.0]{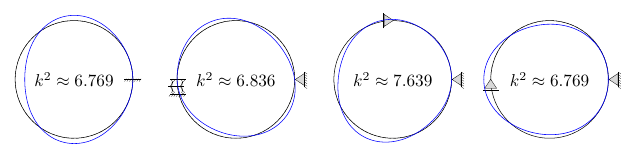}
    \includegraphics[keepaspectratio, scale=1.0]{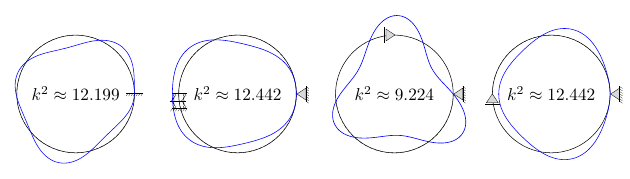}
    \caption{Bifurcation modes under centrally-directed radial load, for different (statically determined) external constraints, in particular, from left to right: one clamp, a (vertically and horizontally)  movable clamp and a pin, a horizontal roller and a pin, a vertical roller and a pin.
    From the upper to the lower row: 1st to 3rd mode. The smallest bifurcation load is obtained 
    with a vertical roller and a pin (upper row on the right).}
	\label{centrlly_ovetti}
\end{figure}

When subject to centrally-directed load, the ring is in neutral equilibrium only under rigid-body rotations. Consequently, constraints restricting this movement, such as a movable clamp, do not affect bifurcation modes. However, this is not true for rigid-body translations, so that limiting these displacements influences the bifurcation loads and modes.  
It has been shown in Section \ref{ocadura} that the equilibrium configuration of the circular rod is stable and, therefore, the first fully continuous mode of bifurcation can be realized even in the absence of external constraints. 
Generally, the bifurcation is sensitive to external constraints for centrally-directed load, even when these realize a statically-determined system. This is shown in Fig.~\ref{centrlly_ovetti}, where different bifurcation modes are reported (critical values of $k^2$ are also included), corresponding to four constraint systems. 
From left to right, these are one clamp, a (vertically and horizontally) movable clamp plus a pin, a horizontal roller plus a pin, and a vertical roller plus a pin.
The upper row of the figure reports the first bifurcation mode, while the second and third modes are sketched in the central and lower rows.

The figure vividly shows that the lowest bifurcation load, $k^2=9/2$, reported in \cite{boresi1955refinement,bodner1958conservativeness}, corresponds to the fully continuous bifurcation, which can be realized without external constraints, but also with a vertical roller and a pin. Changing the constraints varies the bifurcation loads, so that $k^2 \approx 6.769$ is the first bifurcation mode for movable clamp plus pin, but corresponds to the second mode for clamp and for vertical roller plus pin. 
The loads $k^2 \approx 5.356$ and $k^2 \approx 6.472$ do not correspond to any higher bifurcation mode occurring for other constraint configurations. 

\subsubsection{(iii) Dead load}

For the dead load, the fully continuous solution (\ref{eq:genConst2}) contains the two rigid-body displacement components, coefficients $a_2$ and $a_3$. 
The structure has to be externally constrained, because otherwise, the dead load would make the structure unstable to rigid-body rotations, Section \ref{ocadura}. 

\begin{figure}[hbt!]
    \centering
    \includegraphics[keepaspectratio, scale=0.95]{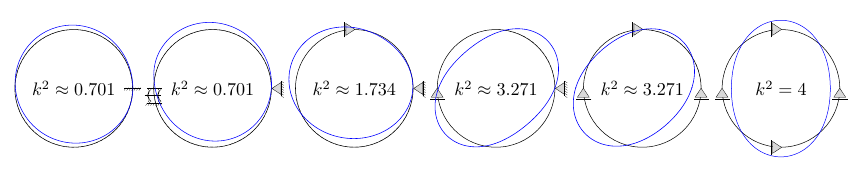}
    \includegraphics[keepaspectratio, scale=0.95]{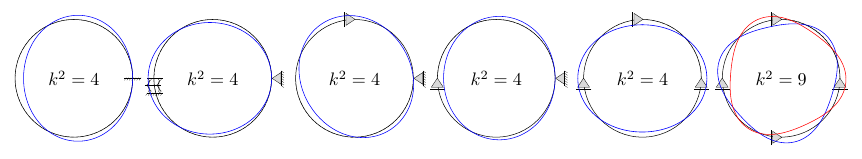}
    \includegraphics[keepaspectratio, scale=0.95]{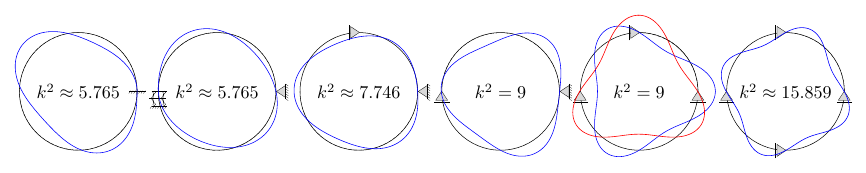}
    \caption{Bifurcation modes under dead radial load, for different (statically determined, plus one undetermined) external constraints, in particular, from left to right: one clamp, a (vertically and horizontally)  movable clamp and a pin, a horizontal roller and a pin, a vertical roller and a pin, 3 rollers, and 4 rollers (statically undetermined). From the upper to the lower row: 1st to 3rd mode. The two modes sketched in red are multiple. The smallest bifurcation load is obtained 
    with one clamp (upper row on the left) or a movable clamp and a pin (upper row, second from the left).}
	\label{rgbm_morto}
\end{figure}

The bifurcation analysis becomes very sensitive to the specific system of external constraints. This is shown in Fig.~\ref{rgbm_morto}, similar to Fig.~\ref{rgbm_constant}, but with a further constraint system where four rollers are used (last column on the right).

The four rollers define a statically undetermined situation, which is included now because in this way the first fully continuous bifurcation mode, $k^2=4$, can be realized. All the other constraint configurations lead to smaller bifurcation loads, initiating with that corresponding to a clamp or a movable clamp plus pin, $k^2\approx 0.701$ (the smallest bifurcation load pointed out in \cite{schmidt1980critical}) and continuing with a roller plus pin and three rollers $k^2\approx 3.271$. Note also that the first fully continuous mode corresponds to the second mode for all constraint systems, except the four rollers. 

As pointed out in \cite{schmidt1980critical}, the bifurcation load $k^2=4$, previously derived by several authors, remains meaningless without a specification of the external constraints applied to prevent rigid-body displacement and rotational instability. Hence, the value reported in \cite{armenakas1963vibrations,bodner1958conservativeness} only refers to the continuous solution and can be obtained by imposing a strong external constraint, as is the case of the four rollers. 
The value $k^2 \approx 3.271$ for roller plus pin constraint was obtained in \cite{schmidt1980critical} to correct the wrong values $k^2\approx 3.265$ provided in \cite{singer1970buckling}.
The fact that there is a bifurcation load $k^2\approx 1.734$, intermediate between $k^2\approx 0.701$ and $k^2\approx 3.271$, passed unnoticed in \cite{schmidt1980critical}.

\section{Experimental set-up for centrally-directed load}

To validate the theoretical results obtained for the bifurcation of a thin ring subject to centrally-directed load and to realize a new type of force distribution never attempted so far, an experimental setup was conceived, designed, realized, and tested in a collaboration between the Laboratory of Integrated Mechanics and Imaging for Testing and Simulations (LIMITS, University of Napoli) and the Instability Lab (University of Trento). 

\begin{figure}[hbt!]
    \centering
    \includegraphics[keepaspectratio, scale=0.12]{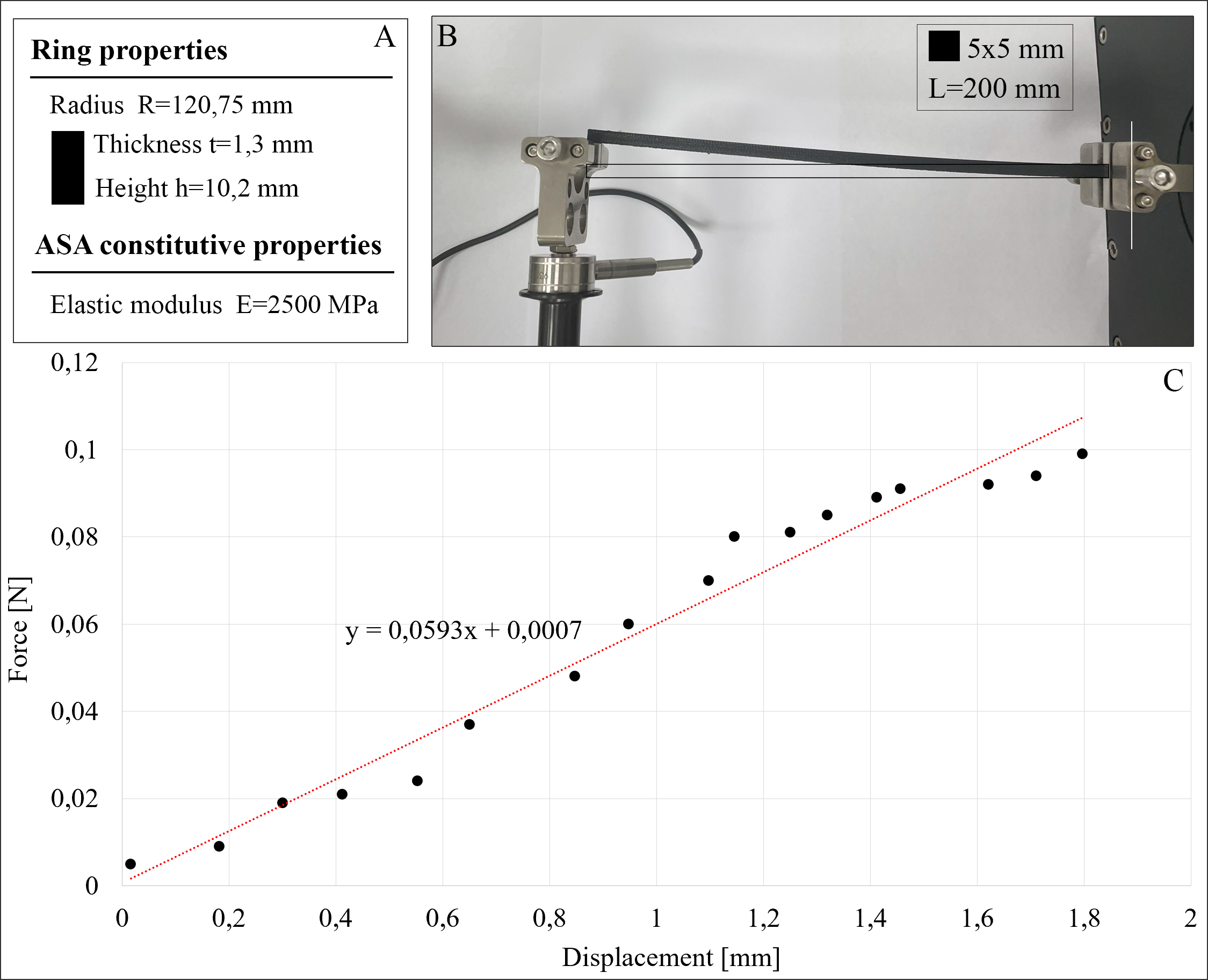}
    \caption{(A) Dimensions of the cross-section of the elastic ring. (B) A photo taken during the bending test performed on a cantilever rod to determine the bending stiffness of the ring and, in turn, to derive its elastic modulus. (C) Experimental points (end force vs. end displacement) recorded during the bending test and showing a remarkable linearity.}
	\label{bending}
\end{figure}

\begin{figure}[hbt!]
    \centering
    \includegraphics[keepaspectratio, scale=0.1]{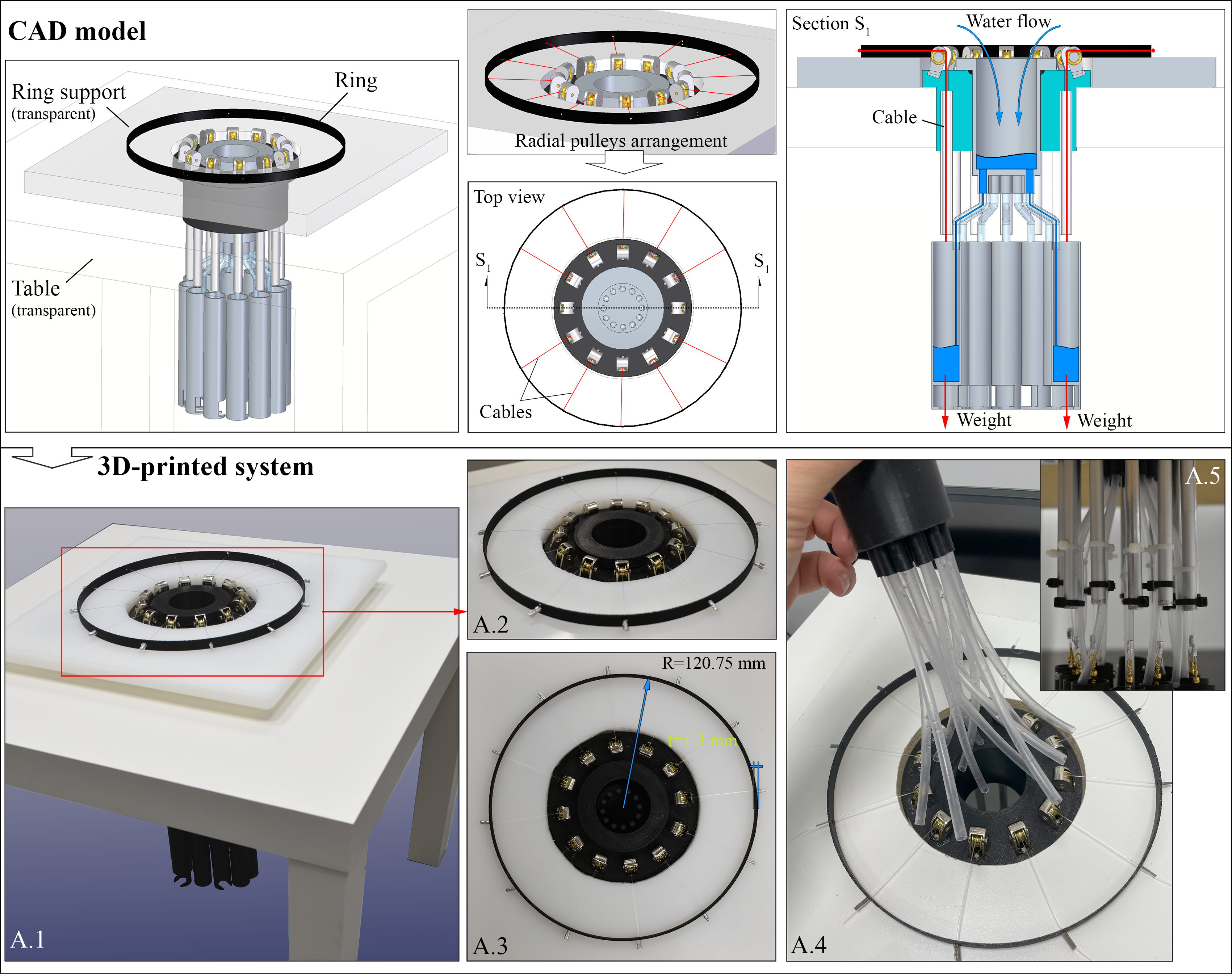}
    \caption{Upper part: CAD virtual model of the conceived experimental set-up, showing the ring (black), the supporting plane (grey), cables (red), pulleys, 
    and the 12 buckets that are filled with water during tests; 
    Lower part, A1-A5: Photos of the experimental set-up, with details showing cable anchorages on the ring and rubber tubes used to fill the buckets with water.}
\label{experiments_setup}
\end{figure}

A ring with radius $120.75$ mm and rectangular ($1.3 \times 10.2$ mm$^2$) cross-section, Fig.~\ref{bending}\,A, was manufactured through 3D printing additive technology (Stratsys Objet 30 Pro), by employing the thermoplastic material \textit{Acrylonitrile Styrene Acrylate} (ASA), a set-up minimizing imperfections, so that possible out-of-roundness have been estimated (through a camera-aided procedure) to be smaller than 10$^{-4}$. 
The elastic stiffness of the material was preliminary measured by manufacturing a rod with a prescribed geometry, to be mechanically tested using the electromechanical machine TA Instruments ElectroForce (200 N 4 motor Planar Biaxial Test Bench) in a cantilever configuration. 
In particular, its Young's modulus, which resulted to be about $2500$ MPa, was determined under bending produced by imposing a dead loading at the free end. The Young modulus was found in agreement with the value declared in the technical datasheet of the material that feeds the 3D printer (see Fig.~\ref{bending}). With another use of additive manufacturing, combined with CAD-based geometry design, components were realized to produce the experimental set-up illustrated in Fig.~\ref{experiments_setup}, which was stabilized by locking it inside a hole made in the central part of a wooden table. To reduce friction effects at the interface between the elastic ring and its support during the experiments, an \textit{ultra-high-molecular-weight polyethylene} (UHMHPE) surface was mounted on the table. 

The centrally-directed load was reproduced by attaching 12 equally spaced cables to the ring. The  number of cables used in the experimental setup was selected based on the results obtained by Albano and Seide for both cases of normal \cite{ albano1973pressure} 
and centrally directed
\cite{albano1973bifurcation}
concentrated forces,  distributed symmetrically along an initially circular rod. They considered the distortion of the configuration due to the discreteness of the loads and analyzed the bifurcation from that state. They showed that, when the loads are at least 5, the average radial load for bifurcation does not differ substantially from that corresponding to the application of a uniform radial load, which leaves the initial configuration undistorted. In particular, for centrally directed radial forces, 12 equally spaced concentrated loads yield a buckling coefficient $k^2=4.505$, almost coincident with the value $k^2=9/2$ corresponding to the radial uniform load. 
The simultaneous application of multiple forces, all of equal intensity, was obtained by designing the device shown in Fig.~\ref{experiments_setup}, where a periodic arrangement of 12 pulleys 
(introduced to minimize friction) allows to convey forces towards the centre of the ring and then downwards through radially-oriented nylon fishing cables ($\phi=0.6$ mm, $F_{max}= 260$ N). The setup ensures that the cables connected to the ring and the pulleys are all lying on the same horizontal plane. 
The centering of the ring and cables was checked with a camera-aided procedure. All parts, including cables, were lubricated with a lithium grease to reduce friction.
The symmetrical distribution of the load among the 12 cables was obtained by pouring water through the central hole at the top of the system, from which the water is channelled and brought to 12 independent buckets, through 12 rubber tubes, progressively filling the tanks.
The geometry of each bucket was sized to initiate tests with a prescribed pre-load still below the instability of the ring (by locating iron weights inside the buckets in a specifically designed housing), then allowing to fill these cylindrical containers up to $40$ gr of water. 

As illustrated in Fig.~\ref{experiments_setup}, the loading process was executed by controlling the amount of water poured with a graduated dosing glass into the buckets. Experiments were recorded during their whole duration, by positioning a camera on the top to follow the different deformation stages of the ring as the applied weight increased, until the the first buckling occurred and the post-buckling initiated.

Two situations were investigated, one in which the ring is left free from external constraints ($k^2=9/2$, bifurcation mode shown in Fig. \ref{hydro_continuity_ovetti}, central part on the left) and the other in which the ring has been constrained with an external clamp ($k^2 \approx  6.472$, bifurcation mode shown in Fig. \ref{centrlly_ovetti}, upper part on the left).
Therefore, two rings with nominally the same characteristics were manufactured and connectors with lobster clasps for each cable were used to reduce manual operations. 
Adopting the set of material and geometrical parameters reported in panel A of Fig.~\ref{bending}, 
from equation \eqref{eq:pcrk}, the expected value of buckling radial load is  $\Pi_{\text{c, r}} \approx 0.0119$ N/mm, corresponding to $k^2=9/2$.

\begin{figure}[hbt!]
    \centering
    \includegraphics[keepaspectratio, scale=0.11]{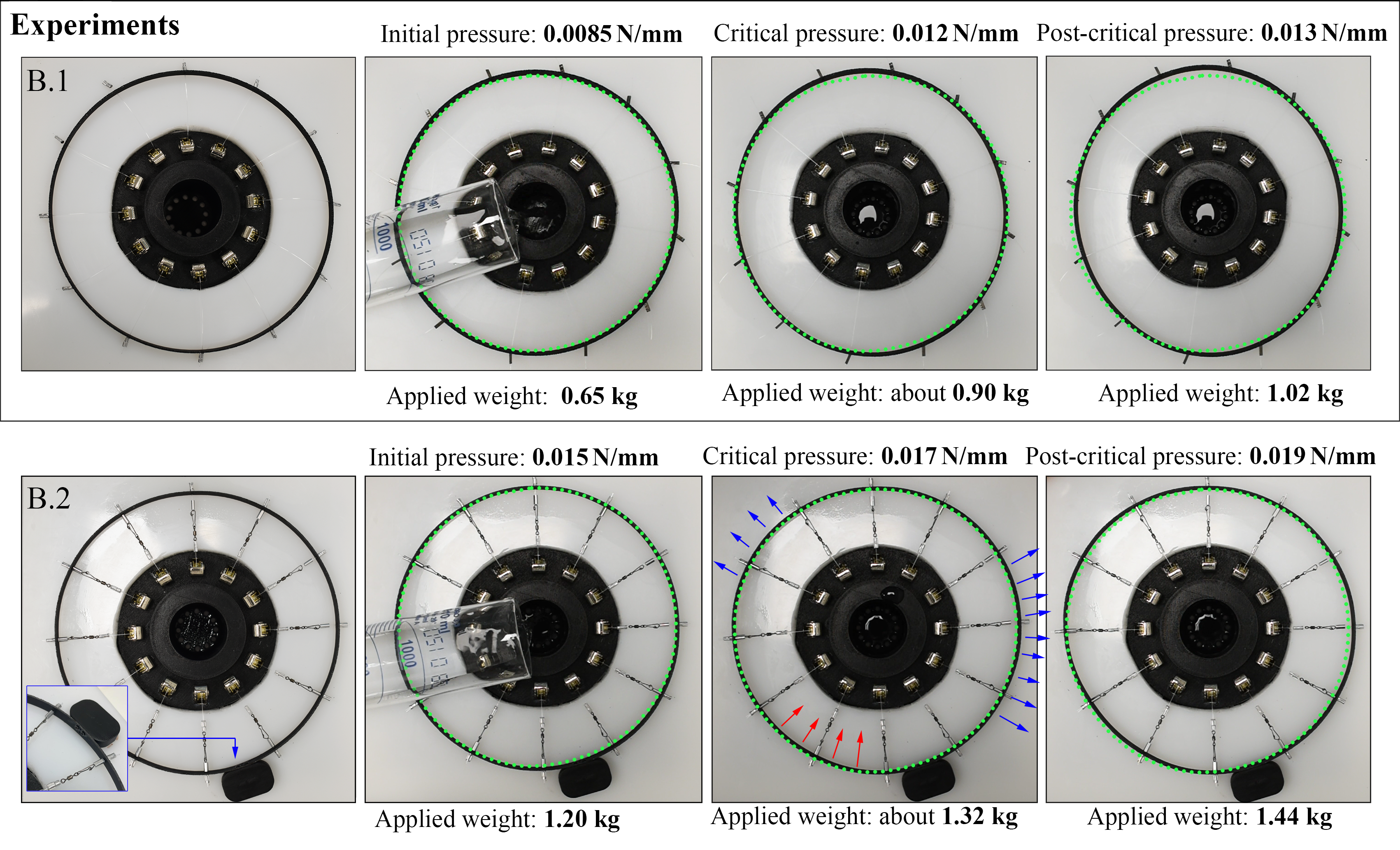}
    \caption{Top views of four instants of the loading process and deformation response of the ring under radially-directed load, without constraints (B.1) and with a point clamped (B.2). From left to right: undeformed configuration of the ring; the graduated dosing glass used to pour water is visible; the ring at a critical pressure (total applied weight $0.90$ kg for ring free from constraints and $1.32$ kg in presence of a clamp), where the green lines highlight the initial circular shape of the undeformed rings against their corresponding ovalization modes; photos of post-critical instants for each of the two cases B.1 and B.2 analyzed at total applied loads of $1.02$ kg and $1.44$ kg, respectively, where the ovalization of the rings become more evident (blue and red arrows in B.2 highlight how the ring tends to protrude and invaginate according to the theoretically predicted first deformation mode).}
\label{experiments}
\end{figure}

Data reported in Fig.~\ref{experiments} (B.1) show that the experiment started from an initial radial load $0.085$ N/mm ($k^2=3.2054$), while bifurcation was found at $0.012$ N/mm ($k^2=4.5253$), and the post-critical behaviour was clearly visible at $0.013$ N/mm ($k^2=4.9024$), where the right panel is in fact representative of the progression of the buckling shape. The experimental results, in terms of both buckling mode (a simple ovalization) and force-equivalent critical radial load ($k^2=4.5253$ instead of $k^2=4.5$), show an excellent agreement with the theoretical predictions, as highlighted by the values reported in Fig.~\ref{experiments}. 
The experimental results confirm that the bifurcation for centrally-directed load, $k^2=9/2$, occurs at a remarkably greater intensity than that for hydrostatic pressure, $k^2=3$, to which a value $\Pi=0.0079$ N/mm for the radial load would correspond. 

Confirmation of theoretical outcomes in comparison with experimental findings, both in terms of critical pressure and (first) deformation mode, were also obtained in the case of the clamped ring, as illustrated in Fig.~\ref{experiments} (B.2). 

From equation \eqref{eq:pcrk}, the expected value of buckling radial load for the ring clamped at a point is $\Pi_{\text{cr}} \approx 0.017$ N/mm, corresponding to $k^2\approx 6.472$. 
For the clamped ring, the experiment started from an initial radial load of $0.015$ N/mm ($k^2=5.6567$), while the bifurcation was found at $0.017$ N/mm ($k^2=6.4109$), and the post-critical behaviour was visible at $0.019$ N/mm ($k^2=7.1651$), the right image in \ref{experiments} (B.2) showing the progression of the ring buckling shape for the case at hand.
The deformed shapes exhibited by the ring at critical and post-critical loads can be compared with the undeformed shape highlighted by the green dotted circles reported in Fig. \ref{experiments}. 

\section{Conclusions}

The bifurcation problem of a circular Euler-Bernoulli rod subject to a uniform radial load is highly sensitive not only to how the load responds to the buckling deformation but, except for the hydrostatic pressure, also to the applied external constraints,  when these define a statically-determined system. Different constraints can, in fact, change the critical load by an order of magnitude for centrally-directed and dead loads. This evidence reconciles previous apparently contradictory statements. 
A new experimental setup demonstrates the feasibility of applying a centrally-directed load to an annular rod. The experiments not only confirm the theoretical predictions but also motivate a new strategy for the design of cable-guided deformable structures.

\section*{Acknowledgements}
The present article is dedicated to Professor Giuseppe, \lq Peppe', Saccomandi who delighted us in several years of sincere friendship, with his enthusiasm, passion for science, and willingness to share ideas in the field of mechanics and beyond.

\vspace{5 mm}
All the authors acknowledge funding from the European Research Council (ERC) under the European Union’s Horizon 2020 research and innovation programme, grant agreement no ERC-ADG-2021-101052956-BEYOND. 
A.C. has also been supported by the Project of National Relevance PRIN2022 grant no P2022XLBLRX and PRIN2022PNRR grant no P2022MXCJ2, funded by the Italian MUR. M.F. additionally thank financial support from MUR through the projects FIT4MEDROB, PNC0000007 (ID 62053) and AMPHYBIA (PRIN-2022ATZCJN).\,

{\newpage
\renewcommand{\bibname}{References}
\bibliographystyle{ieeetr} 
\bibliography{biblio}

\begin{thebibliography}{10}

\bibitem{bresse1859cours}
J.~A.~C. Bresse, {\em Cours de mecanique appliquee: R\'esistance des
  mat\'eriaux et stabilit\'edes constructions}, vol.~1.
\newblock Mallet-Bachelier, 1859.

\bibitem{levy1884memoire}
M.~L{\'e}vy, ``M{\'e}moire sur un nouveau cas int{\'e}grable du probl{\`e}me de
  l'{\'e}lastique et l'une de ses applications,'' {\em Journal de
  Math{\'e}matiques Pures et Appliqu{\'e}es}, vol.~10, pp.~5--42, 1884.

\bibitem{tadjbakhsh1967equilibrium}
I.~Tadjbakhsh and F.~Odeh, ``Equilibrium states of elastic rings,'' {\em
  Journal of mathematical analysis and applications}, vol.~18, no.~1,
  pp.~59--74, 1967.

\bibitem{oran1969buckling}
C.~Oran and R.~S. Reagan, ``Buckling of uniformly compressed circular arches,''
  {\em Journal of the Engineering Mechanics Division}, vol.~95, no.~4,
  pp.~879--895, 1969.

\bibitem{wang1985symmetric}
C.~Wang, ``Symmetric buckling of hinged ring under external pressure,'' {\em
  Journal of Engineering Mechanics}, vol.~111, no.~11, pp.~1423--1427, 1985.

\bibitem{wang1993elastic}
C.~Wang, ``Elastic stability of an externally pressurized ring with two
  hinges,'' {\em ZAMM-Journal of Applied Mathematics and Mechanics/Zeitschrift
  f{\"u}r Angewandte Mathematik und Mechanik}, vol.~73, no.~11, pp.~301--305,
  1993.

\bibitem{kang1994thin}
Y.~J. Kang and C.~H. Yoo, ``Thin-walled curved beams. ii: Analytical solutions
  for buckling of arches,'' {\em Journal of engineering mechanics}, vol.~120,
  no.~10, pp.~2102--2125, 1994.

\bibitem{pi2002plane}
Y.-L. Pi, M.~Bradford, and B.~Uy, ``In-plane stability of arches,'' {\em
  International Journal of Solids and Structures}, vol.~39, no.~1,
  pp.~105--125, 2002.

\bibitem{attard2013plane}
M.~M. Attard, J.~Zhu, and D.~Kellermann, ``In-plane buckling of circular arches
  and rings with shear deformations,'' {\em Archive of Applied Mechanics},
  vol.~83, pp.~1145--1169, 2013.

\bibitem{wangcm}
C.~Wang, H.~Zhang, N.~Challamel, and W.~Pan, {\em Hencky Bar-Chain/Net for
  Structural Analysis}.
\newblock World Scientific, 2020.

\bibitem{giomi}
L.~Giomi and M.~L, ``Minimal surfaces bounded by elastic lines,'' {\em
  Proceedings of the Royal Society A}, vol.~468, pp.~1851--1864, 2012.

\bibitem{catte}
A.~Catte, J.~Patterson, M.~Jones, W.~Jerome, D.~Bashtovyy, Z.~Su, F.~Gu,
  J.~Chen, M.~Aliste, S.~Harvey, L.~Li, G.~Weinstein, and J.~Segrest, ``Novel
  changes in discoidal high density lipoprotein morphology: a molecular
  dynamics study,'' {\em Biophys. J.}, vol.~90, no.~12, pp.~4345--4360, 2006.

\bibitem{savin}
T.~Savin, N.~Kurpios, A.~Shyer, P.~Florescu, H.~Liang, L.~Mahadevan, and
  C.~Tabin, ``On the growth and form of the gut,'' {\em Nature}, vol.~476,
  no.~7358, pp.~57--62, 2011.

\bibitem{napoli}
G.~Napoli and A.~Goriely, ``A tale of two nested elastic rings,'' {\em
  Proceedings of the Royal Society A}, vol.~473, p.~20170340, 2017.

\bibitem{boresi1955refinement}
A.~Boresi, ``A refinement of the theory of buckling of rings under uniform
  pressure,'' {\em Journal of Applied Mechanics}, vol.~22, no.~1, pp.~95--102,
  1955.

\bibitem{bodner1958conservativeness}
S.~R. Bodner, ``On the conservativeness of various distributed force systems,''
  {\em Journal of the Aerospace Sciences}, vol.~25, no.~2, pp.~132--133, 1958.

\bibitem{armenakas1963vibrations}
A.~Armenakas and G.~Herrmann, ``Vibrations of infinitely long cylindrical
  shells under initial stress,'' {\em AIAA journal}, vol.~1, no.~1,
  pp.~100--106, 1963.

\bibitem{singer1970buckling}
J.~Singer and C.~Babcock, ``On the buckling of rings under constant directional
  and centrally directed pressure,'' {\em Journal of Applied Mechanics},
  vol.~37, no.~1, pp.~215--218, 1970.

\bibitem{schmidt1980critical}
R.~Schmidt, ``Critical constant-directional pressure on circular rings and
  hingeless arches,'' {\em Zeitschrift f{\"u}r angewandte Mathematik und Physik
  ZAMP}, vol.~31, pp.~776--779, 1980.

\bibitem{schmidt1981buckling}
R.~Schmidt, ``Buckling of a clamped-hinged circular arch under gas pressure and
  related problems,'' {\em Journal of Applied Mechanics}, vol.~48, no.~2,
  pp.~425--426, 1981.

\bibitem{mascolo2023revisitation}
I.~Mascolo and F.~Guarracino, ``Revisitation of elastic buckling of circular
  rings: Some analytical and numerical issues,'' {\em Thin-Walled Structures},
  vol.~182, p.~110287, 2023.

\bibitem{hoang}
T.~Hoang, ``Influence of chirality on buckling and initial postbuckling of
  inextensible rings subject to central loadings,'' {\em International Journal
  of Solids and Structures}, vol.~172--173, pp.~97--109, 2019.

\bibitem{timoshenko1950strength}
S.~Timoshenko, {\em Strength of Materials: Pt. 1. Elementary Theory and
  Problems}, vol.~1.
\newblock Van Nostrand, 1940.

\bibitem{gaibotti2024bifurcations}
M.~Gaibotti, S.~Mogilevskaya, A.~Piccolroaz, and D.~Bigoni, ``Bifurcations of
  an elastic disc coated with an elastic inextensible rod,'' {\em Proceedings
  of the Royal Society A}, vol.~480, no.~2281, p.~20230491, 2024.

\bibitem{albano1973pressure}
E.~Albano and P.~Seide, ``Bifurcation of circular rings under normal
  concentrated loads,'' {\em Journal of Applied Mechanics}, vol.~40, no.~3,
  pp.~233--238, 1973.

\bibitem{albano1973bifurcation}
E.~Albano and P.~Seide, ``Bifurcation of rings under concentrated centrally
  directed loads,'' {\em Journal of Applied Mechanics}, vol.~40, no.~2,
  pp.~553--558, 1973.

\end{thebibliography}
\nocite{albano1973bifurcation,oran1969buckling}
}

\end{document}